\documentclass[a4paper,11pt]{article}
\usepackage{pos}
\usepackage{physics} 
\usepackage{subcaption}
\usepackage{graphicx}
\usepackage[export]{adjustbox}
\usepackage{array}
\newcolumntype{C}[1]{>{\centering\arraybackslash}p{#1}}
\graphicspath{{./images/}}

\newcommand{\nn}{\nonumber\\}
\newsavebox{\tempfig}

\title{
    Lattice extraction of the Collins-Soper kernel using the auxiliary field 
    representation of the Wilson line
}
\ShortTitle{Lattice extraction of the CS kernel}

\author[a]{Anthony Francis}
\author[a,b]{C.-J. David Lin}
\author*[a]{Wayne Morris}
\author[c]{Yong Zhao}

\affiliation[a]{
    Institute of Physics, 
    National Yang Ming Chiao Tung University,\\
    1001 Ta-Hsueh Road, Hsinchu 30010, Taiwan
}

\affiliation[b]{
    Centre for High Energy Physics, Chung-Yuan Christian University,\\
    Chung-Li 32023, Taiwan
}

\affiliation[c]{
    Physics Division, Argonne National Laboratory,\\
    9700 S. Cass Avenue, Lemont, IL 60439, United States
}

\emailAdd{waynemorris@nycu.edu.tw}

\abstract{
    The Collins-Soper (CS) kernel may be obtained through the TMD soft function 
    by formulating the Wilson line in terms of 1-dimensional auxiliary fermion 
    fields on the lattice. Our computation takes place in the region of the 
    lattice that corresponds to the “spacelike” region in Minkowski space, i.e., 
    Collins' scheme. We explore two methods for obtaining the CS kernel. The 
    ``ratio method''; which would allow us to obtain the soft function as well 
    as the CS kernel. And the ``double ratio''; which allows us to achieve a 
    high degree of statistical precision, but only produces the CS kernel. The 
    matching of our result to Minkowski space is achieved through the mapping of 
    the complex auxiliary field directional vector to the Wilson line rapidity. 
    We present a preliminary extraction of the CS kernel using the 
    ``double ratio'', and discuss the methodology employed.
}

\FullConference{
    The 42nd International Symposium on Lattice Field Theory (LATTICE2025)\\
    2-8 November 2025\\
    Tata Institute of Fundamental Research, Mumbai, India\\
}

\begin{document}
\maketitle

\section{Introduction}

The investigation of transverse momentum dependent (TMD) physics is a central
aspect of mapping the three dimensional structure of hadrons. Well-defined TMD
observables require the regularization of rapidity divergences that appear in
TMD factorization formulas, and give rise to the CS kernel, which governs the
rapidity evolution of TMD observables. The presence of rapidity divergences in 
TMD calculations, however, create a uniqe problem on the lattice, since rapidity 
has no direct anologue in Euclidean space. 
We propose to compute the CS kernel via the same Wilson loop operator
used in the computation of the Soft function, but with directional vectors with 
purely imaginary time components: 
$\tilde n=(in^0,\vec 0_\perp,n^3)$ \cite{Francis:2023, Morris:2025y}. 

Representing the Wilson line in this way allows for a direct analytic 
continuation of the Euclidean soft function to its Minkowski space counterpart 
in the space-like regime, since the soft function is time-independent. 
We model the Wilson line on the lattice via the auxiliary field 
representation, whose equation of motion can be solved iteratively in 
Euclidean time.
There are important issues that arise in the lattice 
realization of an auxiliary field propagator that will be discussed later.
  
There is a direct connection between this 
complex Wilson line direction on the Euclidean lattice and the Minkowski space 
rapidity defined in Collins regularization scheme \cite{collins2011a}, which we
have verified to one-loop in perturbation theory.
In other words, our lattice computation corresponds to the Minkowski space 
soft function defined with space-like directed Wilson lines.

When the Wilson line is pointing in a time-like direction, its auxiliary field 
representation corresponds to heavy quark effective theory (HQET). 
In this case, the directional vector of the Wilson line is the same as the 
heavy quark velocity. 
It was proposed in \cite{jiliuliu2020} to calculate the soft function using 
HQET, modeling the soft function as the form factor of a heavy quark pair. 
In this formulation, the heavy quark velocity corresponds to the time-like 
directed Wilson lines in Minkowski space.
While our work is motivated by this proposal, it differs in two ways: 
our use of Euclidean directional vectors that map to space-like directions in 
Minkowski space, and our method for handling UV cutoff effects. 
Furthermore, we find that a perturbative Euclidean space computation of the soft 
function with directional vectors corresponding to time-like directions in 
Minkowski space gives a divergent integral.

\section{Theoretical considerations}

\begin{figure}[b]
    \centering
    \begin{center}
        \includegraphics[width=0.4\textwidth]{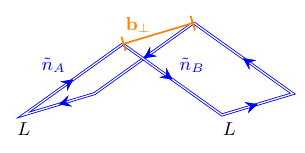}
    \end{center}
    \caption{Butterfly loop computed in Euclidean space.}\label{fig:bfly}
\end{figure}

We have performed a one-loop perturbative computation of the butterfly loop,
Fig. \ref{fig:bfly}, with
infinite and finite length Wilson lines. The infinite length result establishes
an equivalence between the Euclidean and Minkowski space one-loop computations.
Furthermore, the finite length result can be combined into a ratio that becomes
independent of Wilson line length, $L$, as $L$ becomes large.

\subsection{Infinite Wilson lines}
  \begin{figure}
    \centering
    \begin{subfigure}{0.2\textwidth}
      \includegraphics[width=\textwidth]{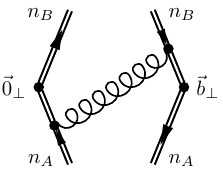}
      \caption{}
      \label{fig:Sa}
    \end{subfigure}
    \begin{subfigure}{0.2\textwidth}
      \includegraphics[width=\textwidth]{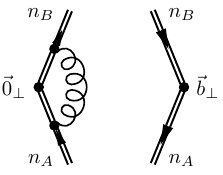}
      \caption{}
      \label{fig:Sb}
    \end{subfigure}
    \begin{subfigure}{0.2\textwidth}
      \includegraphics[width=\textwidth]{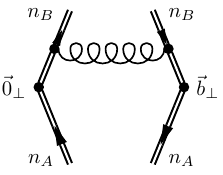}
      \caption{}
      \label{fig:Sc}
    \end{subfigure}
    \begin{subfigure}{0.2\textwidth}
      \includegraphics[width=\textwidth]{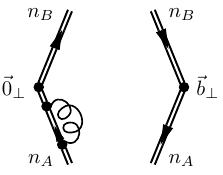}
      \caption{}
      \label{fig:Sd}
    \end{subfigure}
    \caption{Diagrams contributing at one-loop to the soft function, up to mirror diagrams.}
    \label{fig:S}
  \end{figure}

To begin, we define the Euclidean space directional vectors of the Wilson lines
as:
\begin{align} 
    \tilde n_A \equiv (in_A^0, \vec 0_\perp, n_A^3), 
    \qquad \tilde n_B
    \equiv (in_B^0, \vec 0_\perp, -n_B^3) \, . 
    \label{eq:euclvec}
\end{align} 
We also define: $r_a \equiv n_A^3/n_A^0$, and $r_b \equiv n_B^3/n_B^0$ for
convenience.
We use dimensional regularization to handle UV divergences.

Looking at diagram (a) in Fig. \ref{fig:S} as an example, we perform the 
integration using the Schwinger parametrization and complete the square in $k$:
\begin{align}
    S_{\ref{fig:Sa}}\left(b_\perp, \epsilon, r_a, r_b\right)
    &= g^2 C_F  (\tilde n_A \cdot \tilde n_B) \int_{-\infty}^0 \dd s 
    \int_{-\infty}^0 \dd t \int_0^\infty \dd u
    \int \frac{\dd^d k}{(2\pi)^d} 
    e^{-uk^2} 
    e^{-(b_\perp+s\tilde n_A-t\tilde n_B)^2/4u} \nn
    &= \frac{g^2 C_F}{4\pi^{2-\epsilon}}   
    \frac{ \left(b_\perp^2\right)^{\epsilon }\Gamma(1-\epsilon)}
        {2 \epsilon_{\rm IR} }
    \frac 12 
    \log \left(
        \frac{\left(r_a-1\right) \left(r_b-1\right)}
            {\left(r_a+1\right) \left(r_b+1\right)} 
    \right)
    \frac{r_a r_b+1}{ \left(r_a+r_b\right)} \, .
    \label{eq:s1ab}
\end{align}
We obtain a finite result for diagram \ref{fig:Sa} when $|r_a|,|r_b|>1$. This 
can be seen from expanding the term in the exponential of line 1 in 
Eq. (\ref{eq:s1ab}):
\begin{align}
    (b_\perp+s\tilde n_A-t\tilde n_B)^2
    &= b_\perp^2 + s^2 (n_A^0)^2\left( r_a^2 -1 \right)
      + t^2 (n_B^0)^2\left( r_b^2 -1 \right)
      + 2st n_A^0 n_B^0 \left( r_a r_b +1 \right) 
      \, .
      \label{rabIneq}
\end{align}
Equation (\ref{rabIneq}) must be positive in order for the integral in 
Eq. (\ref{eq:s1ab}) to converge. 
Because $s$ and $t$ can vary independently of each other, the terms on the RHS 
of Eq. (\ref{rabIneq}) must separately be greater than zero.

The full one-loop result in Euclidean space with complex directional vectors 
is:
\begin{align}
    S(b_T,\epsilon,r_a,r_b)
    &= 1 + \frac{\alpha_s C_F}{2\pi}    
    \left(\frac{1}{\epsilon } + \log (\pi b_\perp^2 \mu_0^2 e^{\gamma_E})\right)
    \left\{
      2
      +
      \log
      \left(
        \frac{\left(r_a-1\right)\left(r_b-1\right)}
          {\left(r_a+1\right)\left(r_b+1\right)}
      \right)
      \frac{r_a r_b+1}{r_a + r_b}
    \right\} \, .
    \label{eq:ole}
\end{align}

In Minkowski space, using off light-cone directional vectors to regulate the 
rapidity divergence leads to two possible choices: space-like 
(Collins' scheme) or time-like Wilson lines:
\begin{align}
  \text{Time-like:}& \quad 
    n_A = \left( 1+e^{-2y_A}, \vec 0_\perp, 1-e^{-2y_A}  \right)
    , \quad 
    n_B = \left( 1+e^{2y_B}, \vec 0_\perp, -1+e^{2y_B}  \right) \\
  \text{Space-like:}& \quad 
    n_A = \left( 1-e^{-2y_A}, \vec 0_\perp, 1+e^{-2y_A}  \right)
    , \quad 
    n_B = \left( 1-e^{2y_B}, \vec 0_\perp, -1-e^{2y_B}  \right) \, . 
\end{align}
Since the complex directional vectors in Euclidean space are expressed in terms 
of the components of their Minkowski space counterparts, we define $r_a$ and 
$r_b$ using the components of the space-like or time-like directional vectors. 
For the time-like case, we obtain:
\begin{align}
  r_a = \frac{1-e^{-2y_A}}{1+e^{-2y_A}}
  , \quad 
  r_b = \frac{1-e^{2y_B}}{1+e^{2y_B}} \, ,
\end{align}
where we can see that $-1<r_a,r_b<1$. We immediately find that this fails to 
satisfy the condition set by Eq. (\ref{rabIneq}). As for the space-like case, we 
find:
\begin{align}
  r_a = \frac{1+e^{-2y_A}}{1-e^{-2y_A}}
  , \quad 
  r_b = \frac{1+e^{2y_B}}{1-e^{2y_B}} \, ,
  \label{eq:spaceR}
\end{align}
which satisfies Eq. (\ref{rabIneq}). 
The third condition, set by the last term on the RHS of Eq. (\ref{rabIneq}),
leads to the constraint that $n_A^0 n_B^0 (r_a r_b +1) >0$, which indicates that 
the Wilson lines must be both future pointing or both past pointing, 
corresponding to $e^+e^-$ annihilation or DY type processes. 

Substituting Eq. (\ref{eq:spaceR}) into Eq. (\ref{eq:ole}), we recover the 
Minkowski space result \cite{Ebert:2019okf}:
\begin{align}
  S(b_\perp, \epsilon, y_A, y_B)
  &=
  1 + \frac{\alpha_s C_F}{2\pi}
  \left( \frac 1{\epsilon} 
    + \ln \left(\pi b_\perp^2 \mu_0^2e^{\gamma_E} \right) \right)
  \left\{
    2 -
    2 |y_A - y_B|
    \frac{1+e^{2(y_B-y_A)}}{1-e^{2(y_B-y_A)}}
  \right\}
  + \mathcal O(\alpha_s^2) \, .
\end{align}
Since $r_{a,b}$ directly map to the rapidities $y_{A,B}$, one can, in principle, 
compute the soft function for several values of $r_{a,b}$ and fit the resulting 
plot to obtain the intrinsic soft function, $S_I$ in:
\begin{align*}
    S\left( b_\perp, y_A, y_B, \mu\right) 
    \overset{\substack{y_A\to+\infty \\ y_B\to -\infty}}{=} 
    S_I\left( b_\perp, \mu\right)  
    e^{2\gamma_q\left( b_\perp, \mu\right) (y_A - y_B)} \, ,
\end{align*}
and the CS kernel:
\begin{align}
    \gamma_q(\mu, b_\perp)
    &=
    \lim\limits_{\substack{y_A\to+\infty \\ y_B\to -\infty}}
    \frac 12 \pdv{y_n} 
    \log \left( 
        \frac{S(b_\perp, y_n, y_B, \mu)}{S(b_\perp, y_A, y_n, \mu} 
    \right)  \, .
\end{align}

\subsection{Finite Wilson lines}

Lattice computations are constrained by their finite space-time volume, meaning 
that Wilson lines must have a finite length $L$. To account for this, we also 
perform a one-loop calculation for Wilson lines of finite length.
One consequence of finite-length Wilson lines is the appearance of linear 
divergences in $1/a$, which behave like $b_\perp/a$ and $L/a$. 
These divergences can be eliminated by constructing the double ratio:
\begin{align}
    S_{\rm double}\left(b_{\perp,1},b_{\perp,2},a,r_1,r_2,L\right)
    &=
    \left.
        \frac{ S \left(b_{\perp,1}, a, r_1, r_1, L\right)}
        { S \left(b_{\perp,2}, a, r_1, r_1, L\right)}  
    \middle/
        \frac{ S \left(b_{\perp,1}, a, r_2, r_2, L\right)}
        { S \left(b_{\perp,2}, a, r_2, r_2, L\right)}  
    \right.
    \nn
    & = 
    \exp \left[ 
        \left(
            \gamma_q\left( b_{\perp,1}, a\right) 
                - \gamma_q\left( b_{\perp,2}, a\right)
        \right) 2(y_1-y_2)
    \right]
    \nn & \qquad \qquad  
    + \mathcal{O} \left( 
        \frac{b_{\perp,1}^2-b_{\perp,2}^2}{L^2} 
        \left(\frac{1}{r_1-1} - \frac{1}{r_2-1}\right), 
        a^2, \frac{r_{1,2}-1}{r_{1,2}+1} \right) \, ,
        \label{eq:ratio}
\end{align}
where $a$ is some generic regulator used for the UV divergence. The $b_\perp/a$
divergences are eliminated by the ratio at two different values of $r$, and
the $L/a$ divergences are eliminate by the ratio at two different values of 
$b_\perp$.

We have determined that Eq. (\ref{eq:ratio}) holds to one loop in perturbation 
theory, using a Polyakov regulator for the UV. 
By computing the ratio in Eq. (\ref{eq:ratio}) on the lattice for sufficiently 
large $L$, we expect to obtain a time-independent observable with 
power-suppressed corrections of order $b_\perp^2/L^2$, since $L$ in the lattice 
formulation is proportional to Euclidean time.

A consequence of Eq. (\ref{eq:ratio}) is that we can only obtain the difference
between two values of the CS kernel at different values of $b_\perp$ on the
lattice. In order to obtain the CS kernel, we must then match to the CS kernel
at a perturbative value of $b_\perp$.

\section{Strategy of numerical implementation}

A well-known result \cite{gervaisnevau1979, arefeva1980} is that the Wilson 
line can be expressed in terms of a one-dimensional auxiliary fermion field that 
``travels'' along the path traced by the Wilson line: 
\begin{align}
    P \exp 
      \left\{-ig\int_{s_i}^{s_f} \dd s n^\mu A_\mu(y(s)) \right\}
    &= 
    Z_{\psi}^{-1} \int \mathcal D \psi \mathcal D \bar \psi \,
    \psi \bar\psi
    \exp
    \left\{
        i \int_{s_i}^{s_f} \dd s \bar\psi i n\cdot\partial \psi 
        - g_0 \bar\psi n \cdot A \psi
    \right\} \, , 
\end{align}
whose propagator, $H_n(x-y)$, satisfies the Green function equation: 
\begin{align}
    in\cdot D H_n(x-y) = i\delta^{(4)}(x-y) \, ,
    \label{eq:grM}
\end{align}
where $D=\partial+ig_0 A$ is the covariant derivative.
The Euclidean space counterpart of {Eq. (\ref{eq:grM})} is defined with a 
directional vector with a purely imaginary time component, which can be written 
as $\tilde n = (in^0,\vec n)$ in terms of the Minkowski space vector components. 
After applying a Wick rotation, the Euclidean space equation becomes: 
\begin{align}
    i \tilde n \cdot D_E H_{\tilde n} (x_E-y_E) = \delta^{(4)}(x_E-y_E) \, .
    \label{eq:greenE}
\end{align}
As discussed in \cite{agliettieetal1992, Aglietti:1993hf}, meaningful solutions 
to Eq. (\ref{eq:greenE}) can only be obtained with a UV cutoff. Hence, using the 
lattice spacing as our regulator, we can construct a discretized version of this 
propagator.
The central idea is to express the soft function in terms of lattice-regulated 
auxiliary field propagators, $H_{\tilde n}$. The soft function can be shown to 
possess a well-defined continuum limit.

Based on a one-loop analysis we expect the butterfly loop to behave as:
\begin{align}
    S_{\rm bfly}
    &=
    S\left(b_\perp, a, r_a, r_b, \tau\right)
    \overset{\tau\to\infty}{\sim} 
    e^{2 \pi \tau (r_a+r_b)/a} / \tau^4
    \label{eq:auxexp}
\end{align}
in the large $\tau$ limit, where $\tau$ is now the Euclidean time. Noticing that
the term in Eq. (\ref{eq:auxexp}) is independent of $b_\perp$, we can construct
the single ratio:
\begin{align}
    R_{\rm single}\left(b_{\perp,1}, b_{\perp,2}, a, r_a, r_b, \tau\right) &= 
        \frac{S \left(b_{\perp,1}, a, r_a, r_b, \tau\right)}
        {S \left(b_{\perp,2}, a, r_a, r_b, \tau\right)} \, . 
    \label{eq:rsingle}
\end{align}
Based on our perturbative work, we expect the RHS of Eq. (\ref{eq:auxexp}) to be 
an overall multiplicative factor in the large $\tau$ region; therefore,
$R_{\rm single}$ will not suffer from the cutoff effects introduced by the
Euclidean auxiliary propagator.

From here, we construct the double ratio introduced in Eq. (\ref{eq:ratio}) in
order to remove linear divergences. The double ratio should also handle the
operator renormalization, since the soft function is multiplicatively
renormalizable.

We use the same method employed in \cite{Horgan:2009} to solve the auxiliary
field propagator on the lattice, leading to improved stability relative to the
simpler method in \cite{Mandula:1990fit, mandulaogilvie1992}. Our
implementation of \cite{Horgan:2009} corresponds to the
infinite mass limit.

\subsection{Rapidity renormalization}
\label{sec:rren}

Lattice regularization breaks the $ {\rm O} (4)$ rotation symmetry in Euclidean 
space, leading to the need of renormalisation for the directional vectors in 
Eq. (\ref{eq:euclvec}), or equivalently $r_a$ and $r_b$ \cite{Aglietti:1993hf}.
We do not have a method to directly determine
the renormalized values of $r_a, r_b$, but we may infer the renormalized
rapidity through the relation:
\begin{align}
    2 \left( y_1^{\rm ren}\left(a\right) - y_2^{\rm ren}\left(a\right)\right)
    & =
    \frac{
        \log\left( 
            S_{\rm double}
                \left(
                    b_{\perp,1}^{\rm pert},b_{\perp,2}^{\rm pert},r_1,r_2,a
                \right) 
        \right)
    }{
        \gamma_q \left(b_{\perp,1}^{\rm pert}, \mu\right) 
        - \gamma_q \left(b_{\perp,2}^{\rm pert}, \mu\right)
    } + {\rm p.s.c.} \, ,
\end{align}
where $S_{\rm double}$ is given in Eq. (\ref{eq:ratio}), $y^{\rm ren}$ is
the renormalized rapidity, and the denominator on the RHS has two values of
$\gamma_q$ in the perturbative region. Also, p.s.c. refers to power suppressed
corrections of the type in Eq. (\ref{eq:ratio}). Here, we assume the 
renormalization scale $\mu \sim 1/a$, such that the continuum matching is 
trivial. In any case, we can take advantage of the fact that the difference 
between two values of the CS kernel is a renormalization group invariant 
quantity.

Using the renormalized rapidity, we can then write down a formula for the CS
kernel at any value of $b_\perp$:
\begin{align}
    \gamma_q\left(b_\perp, \mu\right)
    &=
    \gamma_q \left(b_{\perp,2}^{\rm pert}, \mu\right)
    +
    \frac{
        \log\left( 
            S_{\rm double}
                \left(
                    b_{\perp},b_{\perp,2}^{\rm pert},r_1,r_2,a
                \right) 
        \right)
    }{2\left(y_1^{\rm ren}\left(a\right) - y_2^{\rm ren}\left(a\right)\right)} 
    + {\rm p.s.c.}\, .
\end{align}

\section{Numerical results and discussion}

\begin{table}[t]
    \centering
    \begin{tabular}{ |C{0.15\textwidth}||C{0.13\textwidth}|C{0.13\textwidth}|C{0.13\textwidth}|C{0.13\textwidth}|  }
     \hline
     $L^3\times T$ & $24^3\times 48$ & $32^3\times 64$ & $40^3\times 80$ & $48^3\times 96$ \\
     \hline
     $a ({\rm fm})$ & 0.081 & 0.060 & 0.048 & 0.041 \\
     \hline
     $N_{\rm config}$ & 400 & 400 & 250 & 341 \\
     \hline
    \end{tabular}
    \caption{Quenched configurations used for computation, with lattice spacing, 
             $a$, and number of configurations, $N_{\rm config}$.}
    \label{tab:config}
\end{table}

\begin{figure}[t]
    \centering
    \includegraphics[width=0.75\linewidth]{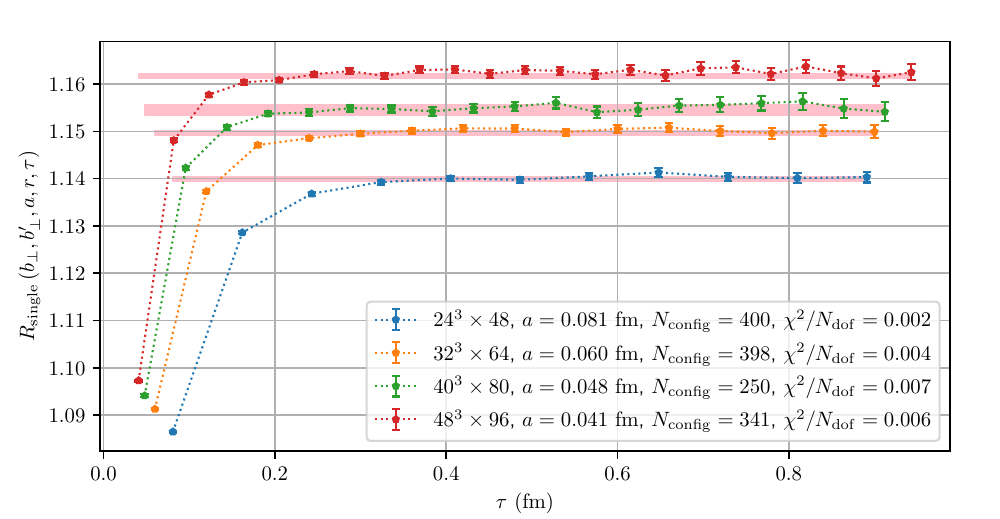}
    \caption{
        Single ratio plot with $b_{\perp} = 4a$, $b'_{\perp} = 3a$, and $r = 
        1.001$
    }
    \label{fig:single}
\end{figure}

For the lattice computation we used the quenched configurations given in
Table \ref{tab:config} \cite{Detmold:2018zgk}. From Fig. \ref{fig:single}, we 
find that the behavior of $R_{\rm single}$ is consistent with our expectations 
from our one-loop analysis; namely, the plateau in Euclidean time suggests that 
UV cutoff effects are cancelled as predicted by Eqs. (\ref{eq:auxexp}) and 
(\ref{eq:rsingle}). We can then perform a constant fit of the plateau to obtain 
$R_{\rm single}$.

Because our procedure requires that we match to perturbative values of the CS
kernel, we need to ensure that we have enough lattice data in the perturbative
region. For our perturbative window we choose 
$3a \leq b_\perp \lesssim 0.2~{\rm fm}$. The lower bound was chosen to avoid
discretization errors from the lattice computation, while the upper bound was
chosen to reduce uncertainty in the perturbative CS kernel. Due to these
restrictions, we omit the $24^3\times 48$ and $32^3\times 64$ lattices from 
the subsequent steps in our analysis procedure.

\begin{figure}
    \centering
    \includegraphics[width=0.85\linewidth]{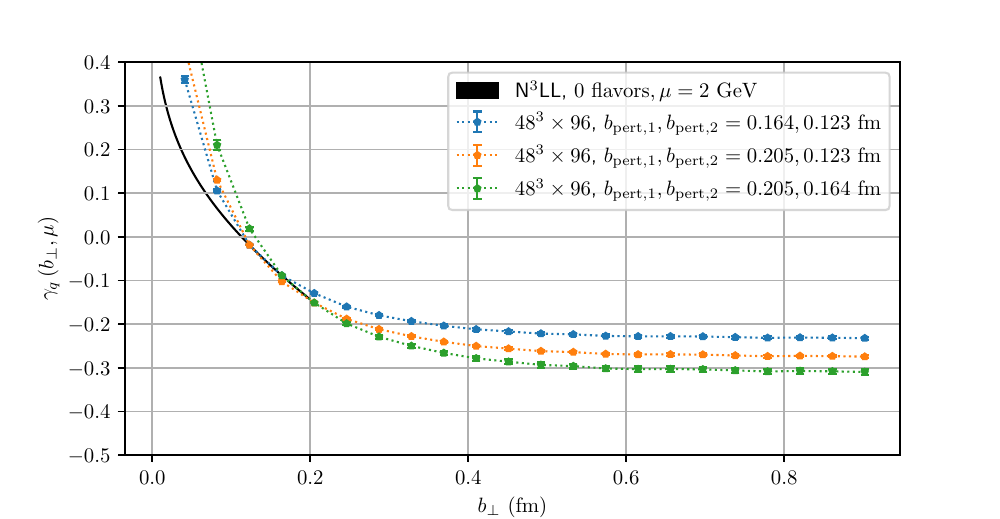}
    \caption{
        CS kernel extracted on the ensemble with $a = 0.041 ~{\rm fm}$ at 
        different $b_{{\rm pert},1}, b_{{\rm pert},2}$ within matching window.
    }
    \label{fig:kcs_stat}
\end{figure}

Following the procedure laid out in Sec. \ref{sec:rren}, we obtain the result
given in Fig. \ref{fig:kcs_stat} for several values of $b_\perp$ within the
perturbative window. We interpret the different results for the CS kernel at
different values of $b_{\rm pert}$ as a systematic uncertainty related to
the use of fixed-order (N$^3$LO) perturbative results.

\begin{figure}[t]
    \centering
    \includegraphics[width=0.85\linewidth]{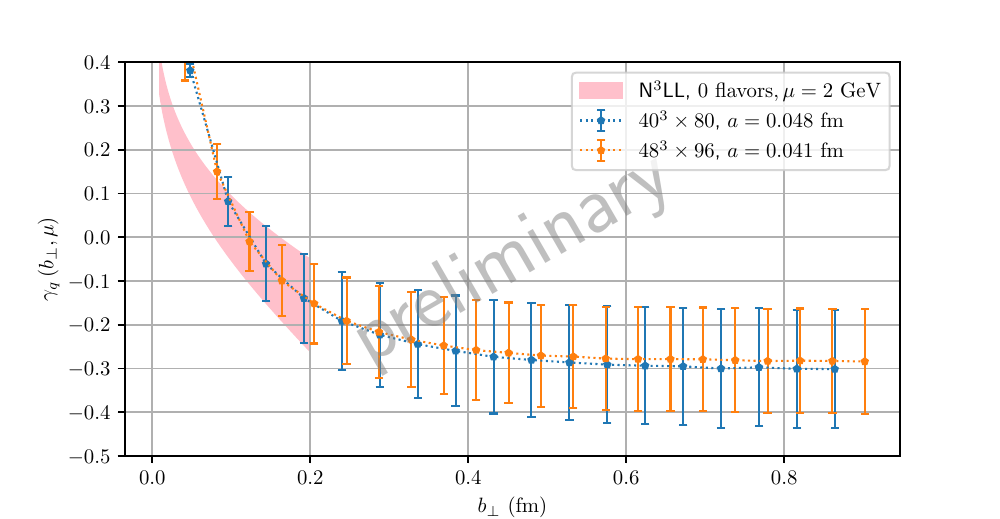}
    \caption{
        The CS kernel extracted by our method, with a conservative
        estimate of systematic error from matching points within 
        fit window and variation of UV scale
    }
    \label{fig:kcs}
\end{figure}

In order to account for the systematic uncertainty, we perform our routine for
extracting the CS kernel for all possible values of $b_{\rm pert}$ within the
matching window. Additionally, we vary the renormalization scale of the
$\overline{\rm MS}$-scheme CS kernel used in the extraction by 10\% around a 
central value of $2~{\rm GeV}$. Performing an average over these results gives 
us a conservative estimate of the systematic error associated with our method, 
as exhibited in Fig. \ref{fig:kcs}.

\section{Conclusions}

We have demonstrated to one-loop in perturbation theory that the auxiliary field
method for extracting the CS kernel from a lattice computation has a strong
theoretical motivation. Particularly, there is a direct mapping between the
Euclidean computation of the butterfly loop and the Minkowski space result at
large lattice time. The double ratio method for computing the CS kernel
difference, 
$\gamma_q\left(b_\perp, \mu\right) - \gamma_q\left(b_\perp', \mu\right)$, allows
us to obtain an high degree of statistical precision. The uncertainty
in our result is dominated by systematic effects, primarily related to the
necessary perturbative matching in our procedure.

We are currently processing data from $64^3\times 128$ quenched configurations
at lattice spacing $a=0.03~{\rm fm}$, which will allow us more control over the
perturbative window. We are also working on a more robust analysis procedure of 
our data to give a better estimate of the systematic uncertainties.

\acknowledgments

    We acknowledge Yizhuang Liu for helpful discussions.
    AF is supported by the National Science and Technology Council (NSTC) of 
    Taiwan under grant 113-2112-M-A49-018-MY3. 
    CJDL and WM are supported by the Taiwanese NSTC through grants 
    112-2112-M-A49-021-MY3, 112-2811-M-A49-517-MY2 and  114-2123-M-A49-001-SVP.
    The work of YZ is supported by the U.S. Department of Energy, Office of 
    Science, Office of Nuclear Physics through Contract No.~DE-AC02-06CH11357 
    and the  Early Career Award through Contract No.~DE-SCL0000017.

\bibliographystyle{JHEP}
\bibliography{bibtex}

\end{document}